\documentclass[aps,pra,reprint, amsmath, amssymb,superscriptaddress]{revtex4-1}

\pdfoutput=1
\usepackage[utf8]{inputenc}
\usepackage{bm}
\usepackage[english]{babel}
\usepackage[T1]{fontenc}
\usepackage{mathrsfs}
\usepackage[retainorgcmds]{IEEEtrantools}
\usepackage{amsmath}
\usepackage{amssymb}
\usepackage{color}
\usepackage{amsfonts}
\usepackage{times,txfonts}
\usepackage{nicefrac}
\usepackage[colorlinks=true,linkcolor=blue,urlcolor=blue,citecolor=blue,pdfusetitle]{hyperref}
\usepackage{BPackage}

\usepackage{float}
\usepackage{epsfig} 
\usepackage{subfigure}
\graphicspath{{Images/}}

\usepackage{physics}
\usepackage{amssymb}
\usepackage[amssymb]{SIunits}

\usepackage{xcolor}

\usepackage{subfiles}
\begin{document}

\title{Entropy production dynamics in quench protocols of a driven-dissipative critical system}
%Options for titles:
%Entropy production dynamics in quench protocols of a driven-dissipative critical system
%Entropy production dynamics of a driven-dissipative phase transition under a sudden quench protocol
%Entropy production dynamics of a driven-dissipative phase transition after a quantum quench
%Dynamics of Wehrl entropy production rate after a quantum quench in a driven-dissipative phase transition
\author{Bruno O. Goes}
\affiliation{Instituto de Física, Universidade de São Paulo, CEP 05314-970, São Paulo, São Paulo, Brazil}

\author{Gabriel T. Landi}
\affiliation{Instituto de Física, Universidade de São Paulo, CEP 05314-970, São Paulo, São Paulo, Brazil}

\begin{abstract}
   Driven-dissipative phase transitions are currently a topic of intense research due to the prospect of experimental realizations in quantum optical setups. The most paradigmatic model presenting such a transition is the Kerr model, which predicts the phenomenon of optical bistability, where the system may relax to two different steady-states for the same driving condition. 
    These states, however, are inherently  out-of-equilibrium and are thus characterized by the continuous production of irreversible entropy, a key quantifier in thermodynamics.
    In this paper we study the dynamics of the entropy production rate in a quench scenario of the Kerr model, where the external pump is abruptly changed. 
    This is accomplished using a  recently developed formalism, based on the Husimi $Q$-function, which is particularly tailored for driven-dissipative and non-Gaussian bosonic systems [Phys. Rev. Res. {\bf 2}, 013136 (2020)].
    Within this framework the entropy production  can be split into two contributions, one being extensive with the drive and describing classical irreversibility, and the other being intensive and directly related to  quantum fluctuations. The latter, in particular, is found to reveal the high degree of non-adiabaticity, for quenches between different metastable states.
\end{abstract}

\maketitle

%%%%%%%%%%%%%%%%%%%%%%%%%%%%%%%%%%%%%%%%%%%%%%%%%
%
%
%               INTRODUCTION
%
%
%%%%%%%%%%%%%%%%%%%%%%%%%%%%%%%%%%%%%%%%%%%%%%%%%

\section{Introduction}
\label{Sec:Intro}

Entropy can be spontaneously generated in a physical process. 
In addition, when a system is connected to an environment, there may also be a flux of entropy. Hence, the entropy of an open system, classical or quantum, evolves according to,
\begin{equation}
\label{eq:Entropy_evolution_OS}
    \frac{dS}{dt} = \Pi - \Phi,
\end{equation}
where $\Pi \geq 0$ is the irreversible entropy production rate and $\Phi$ is the entropy flux from the system to the reservoir. 
Steady states are characterized by $dS/dt = 0$.
These may either be equilibrium steady states (ESSs), characterized by $\Pi_{\text{ESS}} = \Phi_{\text{ESS}} = 0$, or non-equilibrium steady states (NESSs), which occur  in systems connected to multiple baths, or which are externally driven.
In this case $\Pi_{\text{NESS}}=\Phi_{\text{NESS}} > 0$, which means that entropy is being continuously produced within the system, all of which flows to the environment. Entropy production is hence the core concept in quantifying how far from equilibrium a process takes place, or in other words, how irreversible it is.

% There are two types of steady states, where $dS/dt = 0$.
%\textcolor{black}{[Introduzir DPTs, o quanto vem sendo estudado, porque e citar o que já foi feito teorica e experimentalmente]}
A particularly interesting class of non-equilibrium processes are those presenting
driven-dissipative phase transitions (DDPTs)~\cite{hartmann_quantum_2008, diehl_quantum_2008, minganti_spectral_2018}, which occur in quantum optical systems subject to a competition between dissipation and a coherent drive. When non-linear media is involved, these system may undergo a phase transition, where the NESS abruptly changes. This therefore falls in the class of dissipative transitions (also called non-equilibrium  phase transitions in the classical context).  Driven-dissipative transitions can be continuous or
 discontinuous~\cite{marro_dickman_1999,tomadin_nonequilibrium_2011,carmichael_breakdown_2015} and are associated with the closing of the Liouvillian super-operator gap~\cite{kessler_dissipative_2012,minganti_spectral_2018}. 
 They therefore offer the possibility of exploring, in a controlled way, many-body quantum phases without any classical analog.
And for this reason, they have been the focus of significant  theoretical studies~\cite{risken_quantum_1987,kessler_dissipative_2012,lee_unconventional_2013,torre_keldysh_2013,minganti_spectral_2018,sieberer_dynamical_2013,carmichael_breakdown_2015,chan_limit-cycle_2015,mascarenhas_matrix-product-operator_2015,sieberer_nonequilibrium_2014,lee_dissipative_2014,weimer_variational_2015,casteels_power_2016,mendoza-arenas_beyond_2016,jin_cluster_2016,bartolo_exact_2016,casteels_critical_2017,savona_spontaneous_2017,rota_critical_2017,biondi_nonequilibrium_2017,casteels_quantum_2017,raghunandan_high-density_2018,gelhausen_dissipative_2018,foss-feig_emergent_2017,hannukainen_dissipation-driven_2018,gutierrez-jauregui_dissipative_2018,vicentini_critical_2018,hwang_dissipative_2018,vukics_finite-size_2019,patra_driven-dissipative_2019,tangpanitanon_hidden_2019}, as well as experimental investigations~\cite{fink_observation_2017,fitzpatrick_observation_2017,fink_signatures_2018,rodriguez_probing_2017}, in the last decade. 
 
%This paper aims to understand how does the entropy production rate evolves in time for an open quantum system, and how it is related to the NESS the system relaxes to when a non-adiabatic process, implemented here as a \textit{quantum quench} (analogous to an abrupt change of the volume of a gas in classical thermodynamics), is performed. Recently, we proposed a theory of entropy production suited for the class of DDPTs~\cite{goes_quantum_2020}, where these quantities were analysed at the NESS. Hence, in the dynamical scenario, we will have not only the entropy production due to the intrinsic non-equilibrium nature of the system but also a contribution from the non-adiabatic process it was subjected. Recently, the Wehrl entropy production rate formalism was used in closed quantum system presenting a dynamical phase transition in Ref.~\cite{goes_wehrl_2020}, the authors showed that there is a correlation between the maximum of entropy production rate and the critical times where the phase transition occured.

Since the entropy production rate $\Pi$ is the fundamental quantity that characterizes a non-equilibrium process, one may ask the following: How does it behave dynamically, when the system is forced to adjust from one steady state to another? 
A particularly interesting scenario is when the system is subject to a quantum quench, i.e., a sudden change in one of the system's parameters~\cite{calabrese_time_2006}. In thermodynamic terms this is a highly non-adiabatic process, analogous to the free expansion of a gas after a sudden change in volume. 
Hence, in addition to the entropy production due to the intrinsic non-equilibrium nature of the NESS, there will also be a contribution due to this non-adiabatic transient dynamics.
Quench dynamics has been the subject of intense research in the past two decades, although most of the work has been on closed systems undergoing unitary dynamics. 
Examples include the evolution of the expectation values of observables in quantum spin chains~\cite{calabrese_time_2006}, the connection with work statistics~\cite{silva_statistics_2008,dorner_emergent_2012,gambassi_large_2012,varizi_quantum_2020}, and the change in diagonal entropy~\cite{polkovnikov_microscopic_2008,polkovnikov_microscopic_2011,polkovnikov_colloquium_2011,alba_entanglement_2017}. Experimental studies have also been carried out in ultra-cold atoms in optical lattices~\cite{orzel_squeezed_2001,greiner_collapse_2002,greiner_quantum_2002,kinoshita_quantum_2006,bloch_quantum_2012, leonard_monitoring_2017}, where the long coherent times allow one to assess the dynamics over extended time scales.

DDPTs, however, are usually associated with photon loss dissipation, and they are thus inherently open.
As a consequence, both the NESS, as well as the transient dynamics after a quench, should be associated with a finite entropy production. 
Very little is known about the behavior of the entropy production across DDPTs, however, especially concerning the quench dynamics. 
Unfortunately, photon losses are not a standard thermal process, since they effectively occur at zero temperature. As a consequence, the standard formulation of entropy production does not apply~\cite{Timpanaro2019a}.
Recently, we proposed a theory of entropy production suited for photon loss dissipation using concepts from quantum phase space~\cite{santos_wigner_2017}.
This allowed us to incorporate not only thermal, but also vacuum fluctuations. 
This theory is suited for experiments in quantum optical setups, as illustrated in~\cite{brunelli_experimental_2018,Rossi2020}. Originally, it was formulated for Gaussian states and processes. But recently this has been extended for the non-Gaussian case~\cite{goes_quantum_2020}. 
This makes it particularly suited for dealing with DDPTs, specially those involving quenches, where Gaussianity is seldom preserved. 

\begin{figure}[!tbp]
    \centering
    \includegraphics[width = 0.45\textwidth]{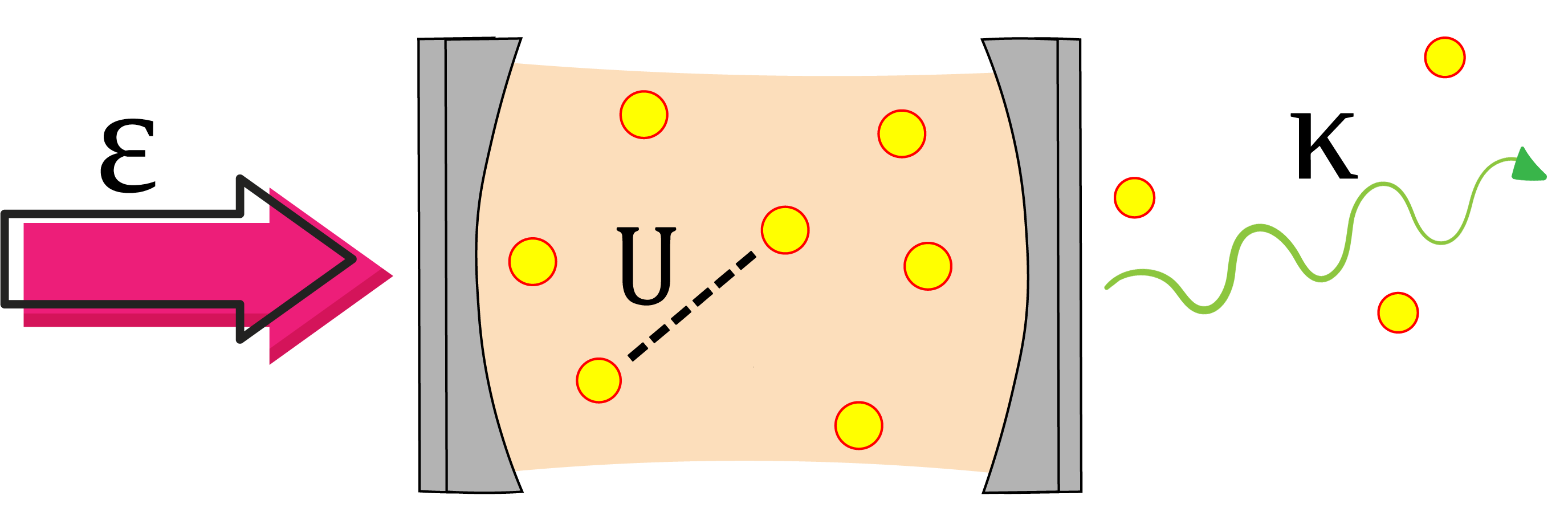}
    \caption{Portrait of an optical cavity with a non-linear medium subject to an external pump $\Epsilon$ and photon losses occurring at a rate $\kappa$, with the effective interaction between photons $U$ as described by the Kerr model.
    }
    \label{fig:KBM_portrait}
\end{figure}

In Ref.~\cite{goes_quantum_2020} this formalism was applied to quantify the entropy production in a NESS. 
The theory, however, is also suited for describing entropy production dynamically. 
This article aims to demonstrate the applicability, and usefulness, of the framework of~\cite{goes_quantum_2020} to study the dynamics of the entropy production in a quench scenario for DDPTs. 
We focus on the paradigmatic Kerr bistability model~\cite{bartolo_exact_2016,casteels_critical_2017,minganti_spectral_2018,roberts_driven-dissipative_2020} which describes an optical cavity filled with a non-linear medium, coherently pumped by a laser, and subjected to an incoherent single-photon loss mechanism (see Fig.~\ref{fig:KBM_portrait}). This model presents a discontinuous DDPT as a function of the external laser drive, which at the mean-field level can be associated with a bistable behavior of the cavity field.
This investigation, as we hope to show, will provide an insightful, and timely, analysis of a problem that lies at the boundaries between statistical physics and quantum optics.

% when a well-defined thermodynamic limit of large photon number is considered. The thermodynamic limit is obtained by letting the driving intensity go to $\Epsilon\rightarrow\infty$ and the non-linearity, responsible for an effective interaction between the photons inside the cavity, goes to $U\rightarrow0$. Their product is kept constant, $\Epsilon U = \text{k}$. 
%

The paper is organized as follows: In Sec.~\ref{Sec:The_KBM} we introduce the Kerr bistability model, its thermodynamic limit, and we briefly discuss the spectral properties of the Liouvillian. In Sec.~\ref{Sec:Wehrl_entropy_prod} we briefly review Ref.~\cite{goes_quantum_2020} and give the expressions for the entropy production/flux rate. 
The formalism is then applied to the sudden quench dynamics scenario in Sec.~\ref{sec:Quench_dyn_ent_prod}. We first present a Gaussianity test, to show the intrinsic non-Gaussian structures that emerge from the quench dynamics. 
This is followed by an exposition of the main results, where we analyze the relevant  thermodynamics during the quench evolution. 
Conclusions are drawn in Sec.~\ref{Sec:conclusions}, where we also discuss possible directions of future research.

%%%%%%%%%%%%%%%%%%%%%%%%%%%%%%%%%%%%%%%%%%%%%%%%%
%
%
%               THE MODEL
%
%
%%%%%%%%%%%%%%%%%%%%%%%%%%%%%%%%%%%%%%%%%%%%%%%%%

\section{Kerr bistability model}
\label{Sec:The_KBM}

\subsection{The Model}
\label{sec:Explaining_the_model}

In this section we review the Kerr bistability model (KBM), which will be the main focus of this work. 
The model is described, \textcolor{black}{in a frame rotating with the pump frequency $\omega_p$ and under rotating-wave approximation \cite{drummond_quantum_1980}}, by the Hamiltonian
\begin{equation}
\label{H}
    H= \Delta\dg{a}a +i\Epsilon(\dg{a}-a)+\frac{U}{2}\dg{a}\dg{a}aa,
\end{equation}
where $\dg{a}$ ($a$) is the single mode creation (annihilation) operator, obeying the bosonic algebra $[a,\dg{a}]=1$. This model describes an optical cavity, filled with a non-linear medium responsible for an effective photon-photon interaction (described by the parameter $U$). The parameter
$\Delta = \omega_c - \omega_p$ is the detuning between the cavity and pump frequencies.
Finally, $\Epsilon$ describes the coherent drive amplitude.
%The environment is the electromagnetic vacuum,i.e., a zero temperature bath.

The system is also subjected to single photon losses at rate $\kappa$ (see Fig.~\ref{fig:KBM_portrait}). 
These  are described within the Born-Markov approximation, so that the state of the system $\rho$ evolves according to the Gorini–Kossakowski–Sudarshan–Lindblad (GKSL) master equation ($\hbar = 1$) \cite{drummond_quantum_1980},
\begin{equation}
\label{M}
    \pd{t}{\rho}=-i[H,\rho]+2\kappa\bigpar{a\rho\dg{a}-\frac{1}{2}\{\dg{a}a,\rho\}} = \mathcal{L}(\rho),
\end{equation}
where $\mathcal{L}$ is the Liouvillian super-operator. 
%The symbol $[A,B]$ represents the commutator between the operators $A$ and $B$ while $\{A,B\}$ stands for the anti-commutator.

%%%%%%%%%%%%%%%%%%%%%%%%%%%%%%%%%%%%%%%%%%%%%%%%%
%
%         THERMO LIMIT AND BISTABILITY
%
%%%%%%%%%%%%%%%%%%%%%%%%%%%%%%%%%%%%%%%%%%%%%%%%%

\subsection{Thermodynamic limit}
\label{sec:Thermo_limit}

The system described by Eq.~\eqref{M} can undergo a phase transition for sufficiently large  pump intensities $\Epsilon$. 
To clarify the nature of this transition, and also better connect it to the standard statistical mechanics literature, it is convenient to define a fictitious thermodynamic limit, which is where the transition actually takes place.
That is, we introduce a dimensionless parameter $N$ and define the thermodynamic limit to correspond to $N\to\infty$. 
This parameter is introduced in a reparametrization of the pump intensity, as   $\Epsilon=\sqrt{N}\epsilon$, where $\epsilon$ is finite. 
The first moment is known to behave as 
$\langle a \rangle \propto \Epsilon$.
We therefore define $\langle a \rangle:=\mu=\sqrt{N}\alpha$, where $\alpha$ is finite and represents the order parameter of the system. 
These parametrizations show that the pump term is extensive, i.e.  $\Epsilon(\mu - \bar{\mu}) \propto O(N)$. 
To properly define the thermodynamic limit, we take as a physical assumption, that this should also hold for the average energy in general. That is,  $\mean{H}\sim O(N)$.
Hence, we must reparametrize all other parameters accordingly. 
In the case of~\eqref{H}, this amounts solely to a rescaling of $U$. 
Since $\mean{\dg{a}\dg{a}aa}\propto O(N^2)$, we therefore reparametrize $U=u/N$~\cite{goes_quantum_2020}.

To summarize, we introduce a parameter $N$ by rescaling  $\Epsilon=\sqrt{N}\epsilon$ and $U=u/N$. 
{\color{black}
This ensures that the relative contribution of the drive and interaction terms remain at similar levels.}
That is, both $\epsilon$ and $u$ are now taken to be finite and the thermodynamic limit is defined as $N\to \infty$.

%We introduce the dimensionless parameter $N$ to have a theoretical control over the thermodynamic limit.  which is defined by $N\rightarrow\infty$ \cite{goes_quantum_2020}. Such a limit is obtained when we have a large pump amplitude $\Epsilon$, hence we parametrize $\Epsilon=\sqrt{N}\epsilon$, where $\epsilon$ is finite. Once $\mean{a} = \mu \propto \Epsilon$, we also parametrize $\mu=\sqrt{N}\alpha$, here $\alpha$ is finite and stands for the order parameter of the system. Finally, the interaction term scales as $\mean{\dg{a}\dg{a}aa}\propto O(N^2)$, leading us to parametrize $U=u/N$, what ensures that the Hamiltonian is extensive in $N$, $H \propto O(N)$.

\subsection{Mean-field behavior and exact solution for the steady-state}

\begin{figure}[t]
    \centering
    \includegraphics[width=0.45\textwidth]{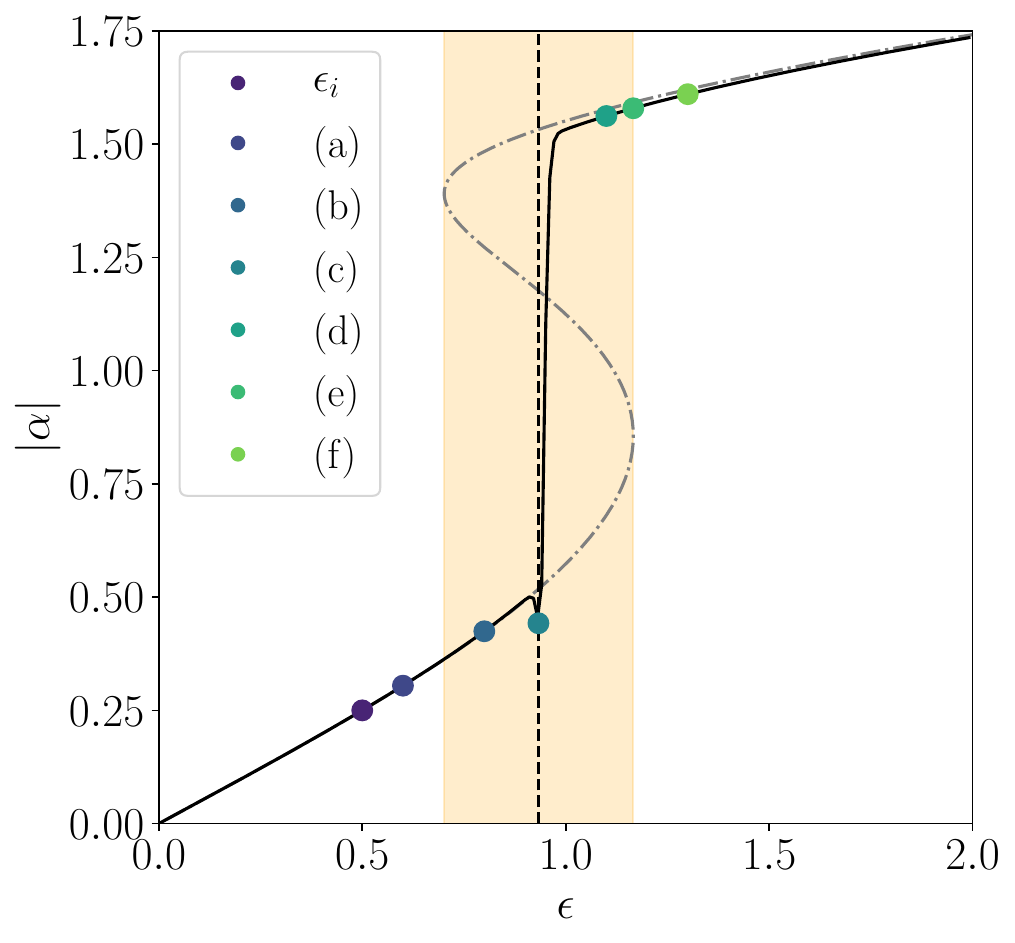}
    \caption{NESS expectation value of the first moment of the Kerr model~\eqref{M}, as a function of the rescaled pump $\epsilon$. The mean-field solution~\eqref{eq:KBM_MFA_sol} is shown by gray dot-dashed lines, while the exact solution~\eqref{eq:KBM-Exact_moments} is shown in black (for $N = 20$). 
    We also mark in the figure the points $\epsilon_i = 0.5$ and $\epsilon_f$, of the quench protocol, which will be used in Sec.~\eqref{sec:Quench_dyn_ent_prod}; viz., (a)-(f) $\epsilon_f = 0.6$, $0.8$,  $\epsilon_c$,  $1.1$, $\epsilon_+$, $1.3$, where $\epsilon_c = 0.933$ is the critical pump and $\epsilon_+ = 1.16616$ marks the edge of the bistability region.
    Other parameters were fixed at $\kappa = 1/2$, $\Delta = -2$, and $u = 1$.
    }
    \label{fig:firstmoment_with_plotmarkers}
\end{figure}

Within the mean-field approximation, one sets $\langle a^\dagger a a \rangle \simeq N^{3/2} |\alpha|^2 \alpha$. 
As a consequence, the equation of motion for the scaled amplitude $\alpha$, becomes
\begin{equation}
\label{eq:KBM_amp_dyn_Rescaled_MFA}
\pd{t}\alpha = -(\kappa +i\Delta+iu |\alpha|^2)\alpha+ \epsilon,
\end{equation}
When $\Delta<-\sqrt{3}\kappa$, the steady-state of this equation is known to present a bistable behavior~\cite{drummond_quantum_1980}.
This is illustrated in Fig.~(\ref{fig:firstmoment_with_plotmarkers}).
The bistability region occurs for $\epsilon$ within the interval 
\begin{equation}
\label{eq:KBM_MFA_sol}
\epsilon_\pm = \sqrt{n_\pm \big[\kappa^2 + (\Delta + n_\pm u)^2\big]}, \quad n_{\pm} = \frac{-2\Delta \pm \sqrt{\Delta^2 - 3\kappa^2}}{3u}.
\end{equation}
For $\epsilon_- < \epsilon < \epsilon_+$,  Eq.~\eqref{eq:KBM_amp_dyn_Rescaled_MFA} presents three steady-state solutions, two of which are stable. 
%
%where $n_{\pm} = |\alpha|^2_{\pm}$. 

At the level of the full master equation~\eqref{M}, however, the nature of this bistability is somewhat subtle. 
This is because, as shown in \cite{drummond_quantum_1980} (see also~\cite{vogel_quasiprobability_1989,Habraken_2012,bartolo_exact_2016,roberts_driven-dissipative_2020}), the steady-state is always unique and can, in  fact, be solved analytically.
Indeed, the moments of the system are found to be given by~\cite{drummond_quantum_1980} 
\begin{equation}
    \label{eq:KBM-Exact_moments}
    \mean{(\dg{a})^na^m} = \sqrt{2}\frac{\cj{\xi}^n\xi^m\Gamma(\cj{x})\Gamma(x)}{\Gamma(\cj{x}+n)\Gamma(x+m)}\frac{_0{F}_2(\cj{x}+n,x+m;|\xi|^2)}{_0{F}_2(\cj{x},x;|\xi|^2)},
\end{equation}
where $\xi = 2\Epsilon/iU$ and $x = 2(i\Delta + \kappa)/iU$. Here $_0{F}_2(a,b;c)$ and $\Gamma$ denote the hyper-geometric and  gamma functions, respectively. 
The predictions of this exact solution are shown as the black-solid lines in Fig.~\eqref{fig:firstmoment_with_plotmarkers}.
As can be seen,  at a certain critical value $\epsilon_c$ the system  jumps abruptly from a state with low photon occupation, to another with high occupation.
Henceforth, we shall refer to the state for $\epsilon < \epsilon_c$ as the dark phase and that for $\epsilon > \epsilon_c$ as the bright phase (there is no closed-form expression for $\epsilon_c$, which has to be computed numerically).

%%%%%%%%%%%%%%%%%%%%%%%%%%%%%%%%%%%%%%%%%%%%%%%%%
%
%           SPECTRAL PROPERTIES 
%
%%%%%%%%%%%%%%%%%%%%%%%%%%%%%%%%%%%%%%%%%%%%%%%%%

\subsection{Spectral properties of the phase transition}
\label{sec:spectral_properties}

This apparent contradiction between the mean-field and exact solutions is resolved by analyzing the full spectrum of the Liouvillian~\eqref{M}, defined by~\cite{kessler_dissipative_2012,minganti_spectral_2018}
\begin{equation}
    \mathcal{L}(\rho_k) = \zeta_k \rho_k,
\end{equation}
where $\rho_k$ are the eigenmatrices associated with the eigenvalue $\zeta_k$ (which may be complex, since $\mathcal{L}$ is not Hermitian). The NESS, in particular, is associated with the zero eigenvalue, $\zeta_0 = 0$,
\begin{equation}
    \mathcal{L}(\rho_{\text{NESS}}) = 0
\end{equation}
It can be shown that, in general, $\Re{\zeta_k} \leqslant 0$ \cite{breuer2002theory}, which ensures stability 
of the dynamics under GKSL evolution. We can thus order the eigenvalues as $0 = |\Re{\zeta_0}| < |\Re{\zeta_1}|< \ldots$. The Liouvillian gap is then defined as $\lambda = |\Re{\zeta_1}|$. 
Physically, it determines the slowest relaxation rate in the long-time limit. 
Or, what is equivalent, the tunneling time $\tau_{\text{tun}}$ between the two branches of optical bistability~\cite{risken_quantum_1987}.

\begin{figure}[t]
    \centering
    \includegraphics[width=0.45\textwidth]{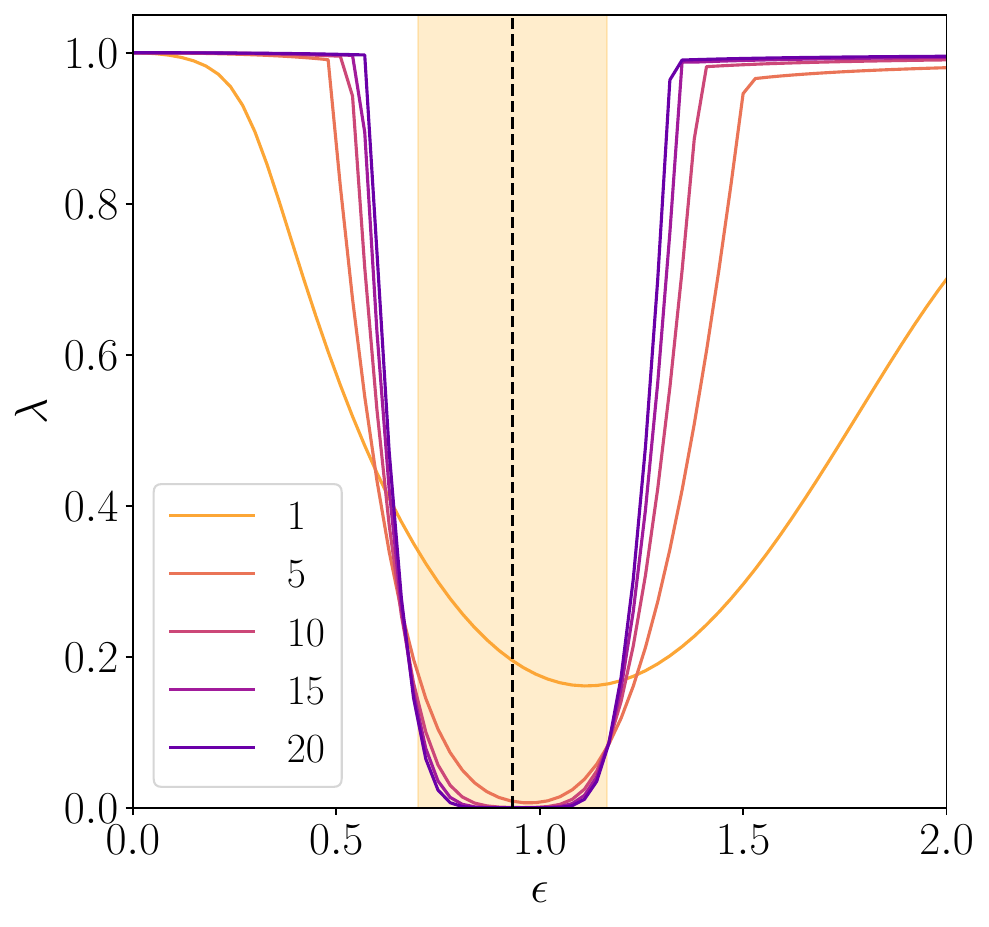}
    \caption{Liouvillian gap $\lambda$ as a function of $\epsilon$.  The dashed black vertical line represents the critical pump $\epsilon_c = 0.933$. \textcolor{black}{Each curve correspond
to a different value of $N$}. The orange patch marks the bistable region, as given by Eq.~\eqref{eq:KBM_MFA_sol}. Other parameters are the same as in Fig.~\ref{fig:firstmoment_with_plotmarkers}.}
    \label{fig:gap_closing}
\end{figure}

As studied in detail in Refs.~\cite{minganti_spectral_2018,kessler_dissipative_2012,casteels_critical_2017}, in the Kerr model the gap closes asymptotically within the entire bistability region $\epsilon \in [\epsilon_-,\epsilon_+]$. 
This is illustrated in Fig.~\ref{fig:gap_closing}.
As a consequence, for any finite $N$, there will always be a unique steady-state, but the gap between this and the first excited state becomes vanishingly small as $N$ increases, thus giving rise to a bistable solution (see also \cite{macieszczak_towards_2016, macieszczak_theory_2020}). 
This hence reconciles the mean-field and exact solutions. 

% The existence of a DDPT is associated with the closing of the Liouvillian gap $\lambda = 0$ in the thermodynamic limit~\cite{minganti_spectral_2018,kessler_dissipative_2012,casteels_critical_2017}. 
% For the Kerr model, however, it turns out that the gap closes asymptotically within the entire bistability region, 
% A detailed study of this was put forth in Ref.~\cite{minganti_spectral_2018}, where the authors studied the spectral properties of general systems undergoing DDPTs. It was found that if one considers $\rho^d$ ($\rho^b$) as the NESS of the dark (bright) phase of the KBM, then one may write the NESS as,
% %
% \begin{equation}
%     \rho_{\text{NESS}} = \theta(\epsilon_c-\epsilon)\rho^d+\theta(\epsilon-\epsilon_c)\rho^b
% \end{equation}
% %
% where $\theta(x)$ is the Heaviside step function.
% %, with $\theta(0) = 1/2$. 
% This means that, for $\epsilon < \epsilon_c$ ($\epsilon > \epsilon_c$) the system relaxes to $\rho^d$ ($\rho^b$).
% %and for $\epsilon = \epsilon_c $ the NESS is a equal mixture of both phases. 
% Moreover, close to the critical point there is a metastable region~\cite{macieszczak_towards_2016, macieszczak_theory_2020} where the system remain in the state it was initialized for a time proportional to $\lambda^{-1}$, which can be infinitely large, before reaching the NESS (see Fig.~\ref{fig:gap_closing}).

%{\color{black}Essa seção do gap ta muito boa e muito bem escrita. Mas faltou concluir; faltou linkar essa discussão toda com a ideia de bistabilidade. Na seção anterior você tava contrastando a mean-field com a solução exata e parecia haver uma incompatibilidade entre as duas. A solução para esse impasse vem exatamente do fato do gap fechar assintóticamente na região de bistabilidade, no limite termodin
%âmico. Precisava incluir isso daí.}\textcolor{purple}{Eu inclui esta discussão ali em cima.}

%%%%%%%%%%%%%%%%%%%%%%%%%%%%%%%%%%%%%%%%%%%%%%%%%
%
%
%           WEHRL ENTROPY PROD FOR KBM
%
%
%%%%%%%%%%%%%%%%%%%%%%%%%%%%%%%%%%%%%%%%%%%%%%%%%

\section{Wehrl entropy production rate for the KBM}
\label{Sec:Wehrl_entropy_prod}

To study the thermodynamics of the KBM model, we use a phase space method based on the Husimi $Q$-function, introduced in Ref.~\cite{goes_quantum_2020}. 
A quantum-phase space formulation of the entropy production  was first introduced in Ref.~\cite{santos_wigner_2017}, and also experimentally assessed in Ref.~\cite{brunelli_experimental_2018}. The problem was formulated in terms of the Wigner function, which generates physical probability densities for Gaussian systems subjected to Gaussian dynamics. This procedure make it possible to circumvent the so called ultra-cold catastrophe~\cite{uzdin2019passivity}; that is, the divergence of the usual entropy production rate and entropy flux rate in the zero temperature limit. 
The Wigner function, however, can become negative, which would lead to complex entropies.
To amend this, one can alternatively use the  Husimi $Q$-function, defined as
\begin{equation}
    Q(\mu,\cj{\mu}) = \frac{1}{\pi}\bra{\mu}\rho\ket{\mu},
\end{equation}
where $\ket{\mu}$ is a coherent state and $\cj{\mu}$ denotes  complex conjugation. The $Q$-function is such that $Q(\mu,\cj{\mu}) \geq0$ \cite{gardiner2004quantum} and can be operationally interpreted as the probability distribution of the outcomes of a heterodyne measurement~\cite{serafini2017quantum} (see also Appendix.~\ref{App:heterodyne_mesurement}).
As the basic entropic quantifier, we use the Wehrl entropy \cite{wehrl_general_1978}, defined as the Shannon entropy of the $Q$-function:
\begin{equation}
\label{eq:Wehrl_entropy_def}
    S(Q) = - \int \dd^2\mu \; Q\ln Q.
\end{equation}
Physically, $S(Q)$ quantifies the entropy of the system, convoluted with additional noise introduced by measuring in the coherent state basis.
As a consequence, $S(Q)$ upper bounds the von Neumann entropy, $S(\rho)=-\Tr{\rho\ln{\rho}}$, i.e. $S(\rho)\leq S(Q)$.  
The Wehrl entropy may thus be viewed as a semi-classical approximation to the von Neumann entropy. 
But despite these limitations, this is currently the only available framework for dealing with the thermodynamics of DDPTs.

We proceed by mapping the master equation Eq.~\eqref{M} into a quantum Fokker-Planck equation
\begin{equation}
\label{eq:QFPeqn}
    \pd{t}Q = \mathcal{U}(Q) + \pd{\mu}J(Q)+\pd{\cj{\mu}}\cj{J}(Q)
\end{equation}
where $\mathcal{U}(Q)$ and $J(Q)$ are differential operators associated with the unitary and dissipative parts of~\eqref{M}. 
The exact form of these quantities can be found using standard correspondence tables \cite{gardiner2004quantum}. 
The latter, in particular, reads 
\begin{equation}\label{J}
    J(Q)= - \kappa(\mu Q + \pd{\cj{\mu}}Q),
\end{equation}
and can be interpreted as the irreversible quasi-probability current associated with the photon loss dissipator.
One may verify that $J(Q)$ vanishes when $Q$ is the vacuum state, which is the fixed point of the dissipator (although not of the whole master equation, due to the presence of the unitary term). 
This corroborates its interpretation as a quasiprobability current~\cite{santos_wigner_2017}.

The unitary term, on the other hand, is somewhat lengthier and reads
\begin{equation}
\label{eq:KBM_Q_unitary_contribution}
\mathcal{U}(Q) = \mathcal{U}_{\Delta} +\mathcal{U}_{\Epsilon} +\mathcal{U}_{U},
\end{equation}
where
\begin{align}
\label{eq:KBM_Q_unitary_contribution_Delta}
\mathcal{U}_{\Delta} &= i \Delta(\mu\pd{\mu}{Q}-\cj{\mu}\pd{\cj{\mu}}{Q}),\\[0.2cm]
\label{eq:KBM_Q_unitaryE}
\mathcal{U}_{\Epsilon} &= -\Epsilon (\pd{\mu}{Q}+\pd{\cj{\mu}}{Q}),\\[0.2cm]
\label{eq:KBM_Q_unitaryU}
\mathcal{U}_{U}&= \frac{i U}{2}[2|\mu|^2(\mu \pd{\mu}{Q}-\cj{\mu}\pd{\cj{\mu}}{Q})+\mu^2\dpd{\mu}{Q}-\cj{\mu}^2\dpd{\cj{\mu}}{Q}].
\end{align}
One notices, in particular, that the photon-photon interaction term, being quartic, leads to terms in $\mathcal{U}_U$ containing second-order derivatives. 
This is a special feature of quantum non-Gaussian models. In classical Fokker-Planck equations, second-order derivatives stem only from dissipation. 
In quantum models, on the other hand, one may in general obtain derivatives of arbitrary orders.

Differentiating \eqref{eq:Wehrl_entropy_def} with respect to time, we find 
\begin{equation}
\label{eq:Wehrl_rate}
   \frac{dS}{dt} = - \int \dd^2\mu \; (\pd{t}Q)\ln Q.
\end{equation}
Inserting~\eqref{eq:QFPeqn} in \eqref{eq:Wehrl_rate},
we can then split $dS/dt$ as [c.f.~Eq.~\eqref{eq:Entropy_evolution_OS}]
\begin{equation}
\label{eq:Wehrl_rate_comparison}
\frac{dS}{dt} = \Pi_{\mathcal{U}} + \Pi_J - \Phi,
\end{equation}
% \begin{align}
% \label{eq:Wehrl_rate_comparison}
%   \frac{dS_t}{dt} 
%   &= - \int \dd^2\mu\;\mathcal{U}(Q)\ln Q- \int \dd^2\mu\;(\pd{\mu}J(Q)+\pd{\cj{\mu}}\cj{J}(Q))\ln Q \nonumber \\
%     &= \Pi_t - \Phi_t
% \end{align}
% %
where
\begin{align}
    % \label{eq:Pi_identification}
    % \Pi_t &\doteq\Pi_{\mathcal{U}}+\Pi_J,\\
    \label{eq:Pi_u_def}
    \Pi_{\mathcal{U}} &:= -\int \dd^2\mu\; \mathcal{U}(Q)\ln Q,\\
    \label{eq:dissipative_contribution_id}
    \Pi_J - \Phi  &:= -
    \int \dd^2\mu\;\Big\{\pd{\mu}J(Q)+\pd{\cj{\mu}}\cj{J}(Q)\Big\}\ln Q.
    % \int \dd^2\mu\; \mathcal{D}(Q)\ln Q.
\end{align}
The term $\Pi_{\mathcal{U}}$ represents how the unitary term contributes to $dS/dt$. 
This is a feature which is unique of phase-space entropies, such as~\eqref{eq:Wehrl_entropy_def}, and is associated with their coarse-grained nature. 
It will be discussed further below. 
The terms $\Pi_J$ and $\Phi$, on the other hand, refer to the splitting of the dissipative contribution into a term associated with an entropy production rate, $\Pi_J$, and another identified as an entropy flux rate $\Phi$.

% A detailed analysis of the two terms in  Eqs.~\eqref{eq:Pi_u_def} and~\eqref{eq:dissipative_contribution_id} was made in~\cite{goes_quantum_2020}.
% In order to make the paper more self-contained, we now briefly review this analysis within the specific context of the KBM model~\eqref{M}.

%%%%%%%%%%%%%%%%%%%%%%%%%%%%%%%%%%%%%%%%%%%%%%%%%
%
%                DISSIPATION 
%
%%%%%%%%%%%%%%%%%%%%%%%%%%%%%%%%%%%%%%%%%%%%%%%%%

\subsection{Dissipative contribution}
\label{sec:Dissipative_Pi_Phi}

As shown in~\cite{goes_quantum_2020,santos_wigner_2017}, the correct splitting of $\Pi_J$ and $\Phi$ in Eq.~\eqref{eq:dissipative_contribution_id} should have the form 
% We start with the dissipative term~\eqref{eq:dissipative_contribution_id}.
% In what follows we assume, as usual, that both the Husimi $Q$-function, as well as its derivatives, vanish at infinity, so that boundary terms do not contribute when integrating by parts. 
% This allows us to write 
% \[
% -\int \dd^2\mu\;\Big\{\pd{\mu}J(Q)+\pd{\cj{\mu}}\cj{J}(Q)\Big\}\ln Q  
% =\int \dd^2\mu\;\frac{\pd{\mu}Q}{Q}J(Q)+\frac{\pd{\cj{\mu}}Q}{Q}\cj{J}(Q).
% \]
% To identify which part corresponds to $\Pi_J$ and which corresponds to $\Phi$, we follow a standard procedure, first developed for classical systems (c.f. Refs~\cite{van_den_broeck_three_2010,tome_entropy_2010,tome_entropy_2012,seifert_stochastic_2012}), but also used in the quantum case~\cite{santos_wigner_2017}. 
% The basic idea is to use Eq.~\eqref{J} to substitute for $\partial_\mu Q$ in terms of $J$. 
% This leads to 
\begin{IEEEeqnarray}{rCl}
    \label{eq:entropy_production_rate}
    \Pi_J &=& \frac{2}{\kappa}\int \dd^2\mu\;\frac{|J(Q)|^2}{Q} \geq 0, \\[0.2cm]
    \label{eq:Total_entropy_flux_rate}
    \Phi &=& \int\dd^2\mu\; \Big(\bar{\mu} J(Q) + \mu \bar{J}(Q)\Big) =  2\kappa\mean{\dg{a} a}.
\end{IEEEeqnarray}
% There are multiple justifications for as to why this splitting is physically consistent. See, for instance, Refs.~\cite{seifert_stochastic_2012} or~\cite{santos_wigner_2017}. 
The quantity $J(Q)/Q$ can be interpreted as a phase-space velocity~\cite{seifert_stochastic_2012}, so that $\Pi_J$  is seen to be proportional to the mean squared phase space velocity, $\mean{|J(Q)/Q|^2}$.

The entropy flux, Eq.~\eqref{eq:Total_entropy_flux_rate},
% the step to the 2nd equality used the fact that $\int \dd^2\mu \; |\mu|^2 Q = \langle a a^\dagger\rangle$. 
% As a result, one finds that the entropy flux 
is associated with the photon loss rate. 
To see that, we start from Eq.~\eqref{M} and compute the time-evolution of the cavity occupation $\langle a^\dagger a \rangle$, which reads
\[
\frac{\dd \langle a^\dagger a \rangle}{\dd t} = \Epsilon \langle a^\dagger + a\rangle - 2 \kappa \langle a^\dagger a\rangle. 
\]
The first term describes the effect of the external coherent pump in populating the cavity, while the second describes the loss rate to the environment. 
The entropy flux is thus found to be directly the photon loss rate, corroborating its interpretation as a ``flux''. 

Within the context of DDPTs, it is interesting to analyze the behavior of $\Pi_J$ and $\Phi$ in the thermodynamic limit (Sec.~\ref{sec:Thermo_limit}). 
In this case it is convenient to define the fluctuation operator \textcolor{black}{$\delta a = a -\sqrt{N}\alpha$, which leads to $\mean{\delta \dg{a}\delta a} = \mean{\dg{a}a} - N|\alpha|^2$.} 
Substituting in Eq.~\eqref{eq:Total_entropy_flux_rate} leads to
\begin{equation}
    \label{eq:splitting_entropy_flux_rate}
    \Phi = \Phi_{\text{ext}} + \Phi_q = 2\kappa |\alpha|^2 N + 2\kappa \mean{\delta \dg{a} \delta a},
\end{equation}
where $\Phi_{\text{ext}} \propto O(N)$ is a clearly extensive contribution, while  $\Phi_q$ is the contribution associated with quantum fluctuations. 

A similar splitting can be done for the entropy production rate~\eqref{eq:entropy_production_rate}. We define the displaced phase space variable $\nu = \mu - \alpha\sqrt{N}$, which leads to a splitting of the current as $J = -\sqrt{N}\kappa \alpha Q + J_\nu(Q) $, where $J_\nu(Q) = - \kappa(\nu Q + \pd{\cj{\nu}}Q)$. Substituting this in \eqref{eq:entropy_production_rate}, we then find
\begin{equation}
     \label{eq:Pi_d}
    \Pi_J = \Pi_{\text{ext}}+\Pi_d =2\kappa|\alpha|^2 N +\frac{2}{\kappa}\int \dd^2\nu\;\frac{|J_\nu(Q)|^2}{Q}.
\end{equation}
As can be seen, the first contribution to the entropy production rate is $\Pi_{\text{ext}}= \Phi_{\text{ext}}$.
This means that the part of the entropy flow associated with the first moments $\alpha$ produces an equal amount of entropy \textcolor{black}{(and hence does not affect the system entropy, according to Eq.~\eqref{eq:Entropy_evolution_OS})}. 
The other contribution, $\Pi_d$, concerns only the quantum fluctuations (we do not call it $\Pi_q$ since there will also be a quantum contribution to $\Pi$ from the unitary part, as we now discuss).

%%%%%%%%%%%%%%%%%%%%%%%%%%%%%%%%%%%%%%%%%%%%%%%%%
%
%           UNITARY CONTRIBUTION 
%
%%%%%%%%%%%%%%%%%%%%%%%%%%%%%%%%%%%%%%%%%%%%%%%%%

\subsection{Unitary contribution}
\label{sec:unitary_contribution}

Next we turn to the unitary term $\Pi_{\mathcal{U}}$ in \eqref{eq:Pi_u_def}.  
An interesting feature of the Husimi $Q$-function is that the only unitary terms which contribute to $dS/dt$ are those involving second-derivatives. 
That is, the terms associated with $\Delta a^\dagger a$ and $i \Epsilon(a^\dagger -a)$, Eqs.~\eqref{eq:KBM_Q_unitary_contribution_Delta} and~\eqref{eq:KBM_Q_unitaryE}, vanish. 
 This can be proven integrating by parts multiple times.
%Appendix~\ref{App:unitary_contribution_calculations}).
The same is also true for the first two terms in Eq.~\eqref{eq:KBM_Q_unitaryU}. 
Thus, the only non-trivial contributions come from the last two terms. 
Integrating by parts multiple times, these can be written as
\begin{equation}
\label{eq:Pi_U_contribution}
    \Pi_{\mathcal{U}} =\frac{i U}{2} \int \dd^2\mu\frac{( \mu ^2(\pd{\mu}Q)^2-\cj{\mu}^2 (\pd{\cj{\mu}}(Q))^2)}{Q}.
\end{equation}
% where we came back to the original coherent state variable $\mu$, but we emphasize that this quantity is $N$-independent, which is clear from Eq.~\eqref{eq:U_contribution_N0_to_Pi_2}.
This contribution to the entropy production stems from the second-order derivatives in the quantum Fokker-Planck equation. Usually, second-order derivatives are associated with diffusion. Since, in this case, these derivatives refer to the unitary part, this cannot be standard diffusion, which would not conserve energy. Instead, the structure appearing in the last two terms of Eq.~\eqref{eq:KBM_Q_unitaryU} represents a special type of diffusion, which is compatible with unitarity. This type of phenomena was studied in~\cite{Altland2012,Altland2012a}, in the context of the Dicke model, where the authors showed that it corresponds to a type of decoherence, linked with the coarse-graining introduced by the phase-space representation. 

Unlike $\Pi_J$, which is always non-negative (as expected by the 2nd law), the sign of $\Pi_\mathcal{U}$ is not well defined. 
This, in a sense, is expected since this quantity measures the effect of the unitary dynamics on $S(Q)$. And one cannot expect that all unitary dynamics always lead to an increase in the Wehrl entropy. 
Notwithstanding, we have recently carried out a study of this type of term in the context of dynamical phase transitions in the Lipkin-Meshkov-Glick model~\cite{goes_wehrl_2020}.
There, we found that even though $\Pi_\mathcal{U}$ did not have a well defined sign, this became asymptotically the case in the thermodynamic limit. 
Thus, our expectation is that for critical models, this term should  tend to be non-negative in the thermodynamic limit (although it is not clear whether this could somehow be proved in general). 
 
We have also carried out a detailed analysis of $\Pi_\mathcal{U}$ in the NESS of the KBM model~\cite{goes_quantum_2020}. 
We found that, quite surprisingly, it behaved similarly to what is expected from the \emph{classical} entropy production in across non-equilibrium transitions~\cite{herpich_collective_2018,noa_entropy_2019,crochik_entropy_2005,shim_macroscopic_2016,herpich_universality_2019}.

%%%%%%%%%%%%%%%%%%%%%%%%%%%%%%%%%%%%%%%%%%%%%%%%%
%
%
%           ENTROPY QUENCH DYNAMICS
%
%
%%%%%%%%%%%%%%%%%%%%%%%%%%%%%%%%%%%%%%%%%%%%%%%%%

\section{Quantum entropy production dynamics}
\label{sec:Quench_dyn_ent_prod}

\subsection{Quench protocol and computational details}

We now arrive at the core of our paper, where we probe the dynamics of the different entropic quantities during a quench protocol of the KBM model. 
The quench causes the system to evolve from the initial NESS, with pump $\epsilon_i$, to a final NESS, with pump $\epsilon_f$. Both initial and final states are already non-equilibrium in nature. 
In addition, however, there will also be an entropy production associated with the quench dynamics, a highly non-adiabatic process.

From this point on, all quantities will be given in units of $\kappa = 1/2$.
We fix $\Delta = -2$ and $u = 1$. 
We also start all quenches from $\epsilon_i = 0.5$.
The only free parameters are then $\epsilon_f$ and $N$.
We will make frequent reference to Fig. \ref{fig:firstmoment_with_plotmarkers}, which shows the different values we chose for $\epsilon_f$. 
For these parameters, the  mean-field bistability region occurs between  $\epsilon_- = 0.701373$ and $\epsilon_+ =1.16616$ [c.f. Eq.~\eqref{eq:KBM_MFA_sol}]. 
Moreover, the  critical point (determined numerically) is at $\epsilon_c = 0.933$. 
The system is thus always initialized in the dark phase, $\epsilon_i=0.5 < \epsilon_c$.
We recall the nomenclature we adopted: the dark phase refers to the first branch, where $\epsilon < \epsilon_c$, and the bright phase refers to the upper branch, where $\epsilon > \epsilon_c$.
Moreover, the region between $[\epsilon_-, \epsilon_+]$ is referred to as the bistability region. 

% The situation of interest here is: set the system initially in the dark phase with $\epsilon_i$, how does the system responds thermodynamically to an abrupt change to an $\epsilon_f$ to the same or to a different phase? What is the role of quantum fluctuations $\Pi_d$ compared to the coarse graining introduced by the the measuring scheme $\Pi_\mathcal{U}$? To address these questions we performed the quench protocol as follows:
 
%In this section we present the behavior of quantum entropy production/flux rates when the system suffers a sudden quench protocol, meaning the driving amplitude $\epsilon$ is abruptly changed.

%Later, we present the results of our numerical simulations to the entropic quantities, which demonstrates the usefulness of the Wehrl entropy production formalism in the dynamical scenario where the state is non-gaussian.

%If the dynamics were Gaussian, this would therefore be an enormous simplification, since it would allow for very efficient means to study the entropy production. 
%There are the so called gaussian states for which we are able to obtain the entropic quantities easily since there is a neat analytical form for the Husimi $Q$-function. Some features of the KBM lead to Gaussian states in the thermodynamic limit. If the dynamics were Gaussian, this would therefore be an enormous simplification, since it would allow for very efficient means to study the entropy production. 

%With that in mind, first we study the gaussianity of the state during the dynamics, and we show that unfortunately it is not the case for our system. 
 %
 
All simulations are done by expressing the relevant operators in the Fock basis, using a sufficiently large number of basis states $n_\text{max}$ to ensure convergence. 
Manipulations involving the Liouvillian~\eqref{M} are then carried out using standard vectorization methods~\cite{Vectorization}.
The quench protocol we investigated can be summarized as follows: 
\begin{enumerate}
    \item The system is initialized in the steady-state of the Liouvillian $\mathcal{L}_i$, associated with  $\epsilon = \epsilon_i$. That is, $\rho_i$ is the solution of $\mathcal{L}_i(\rho_i) = 0$.
    
    \item At time $t=0$ we abruptly change the pump amplitude to a certain value $\epsilon_f$, which defines a new Liouvillian $\mathcal{L}_f$;
    
    \item For $t>0$ the state evolution is governed by the final Liouvillian according to 
    \[
    \rho_t=e^{\mathcal{L}_f t}\rho_i,
    \]
    which is  the formal solution of~\eqref{M}.
    $\rho_t$ is computed numerically by discretizing the evolution in small time steps, usually $\Delta t = 0.2$;
    
    \item At each time step we compute $\langle a \rangle$ and $\langle a^\dagger a \rangle$. From the former we find $\alpha = \langle a \rangle/N$ and from the latter we obtain the net entropy flux rate~\eqref{eq:Total_entropy_flux_rate}, $\Phi = 2 \kappa \langle a^\dagger a \rangle$. 
    % The quantum entropy flux rate $\Phi_q$ in Eq.~\eqref{eq:splitting_entropy_flux_rate} is then  easily obtained as $\Phi_q = \Phi - 2\kappa |\alpha|^2 N $;
    
    \item The Husimi $Q$-function is computed numerically at each time step, by constructing approximate coherent states,
    \begin{equation}
        \ket{\mu} = e^{-|\mu|^2/2} \sum\limits_{n=0}^{n_{\text{max}}} \frac{\mu^n}{\sqrt{n!}} \ket{n}.
    \end{equation}
    We compute $Q(\mu, \bar{\mu})$ for a sufficiently fine grid of points ($\mu, \bar{\mu})$ in the complex plane. 
    %
    % the convergence of the numerical evolution is ensured by choosing different truncation sizes $n_{\text{max}}$ for the dimension of the Fock space, depending on the value of $\epsilon_f$; %(see App.~\ref{App:Computational_details});
    
    \item The entropy production rates $\Pi_J$ and $\Pi_\mathcal{U}$ in Eqs.~\eqref{eq:entropy_production_rate} and~\eqref{eq:Pi_U_contribution}  are computed numerically, by integrating the corresponding functions over the complex plane grid. 
    Derivatives of the Husimi function can be computed efficiently using the Bargmann state \cite{gardiner2004quantum}, as detailed in Appendix~\ref{App:Details_Pies}.
\end{enumerate}

These simulations are computationally costly, and the cost increases significantly with $N$.
The reason is twofold. 
First, larger values of $N$ require a larger number of basis elements $n_\text{max}$, to ensure convergence. 
Second, larger $N$ also requires larger grids in the complex plane, which significantly increases the cost of the numerical integrations involved in computing $\Pi_J$ and $\Pi_\mathcal{U}$.

We begin in Sec.~\ref{sec:Gauss_tests} by showing that, in general, the dynamics is not Gaussian.
This is interesting, since in the thermodynamic limit this would be the case for the NESS. 
However, due to the highly non-adiabatic nature of the quench, the intermediate states will in general deviate significantly from Gaussianity. 
Next we turn to the properties of the entropy production rate and the flux rate. 
In Sec.~\ref{Sec:Entropy_flux_dyn} we study the entropy flux rate $\Phi$ in Eq.~\eqref{eq:splitting_entropy_flux_rate} and in Sec.~\ref{Sec:Entropy_flux_dyn} we study the component $\Pi_d$ [of $\Pi_J$, Eq.~\eqref{eq:Pi_d}], and the unitary contribution $\Pi_\mathcal{U}$ in~\eqref{eq:Pi_U_contribution}.

 %%%%%%%%%%%%%%%%%%%%%%%%%%%%%%%%%%%%%
 %
 %      GAUSSIANITY TESTS
 %
 %%%%%%%%%%%%%%%%%%%%%%%%%%%%%%%%%%%%%
 
 \subsection{Gaussianity of the state during the dynamics}
 \label{sec:Gauss_tests}
 
\begin{figure}[htp]
\centering
\includegraphics[width=0.45\textwidth]{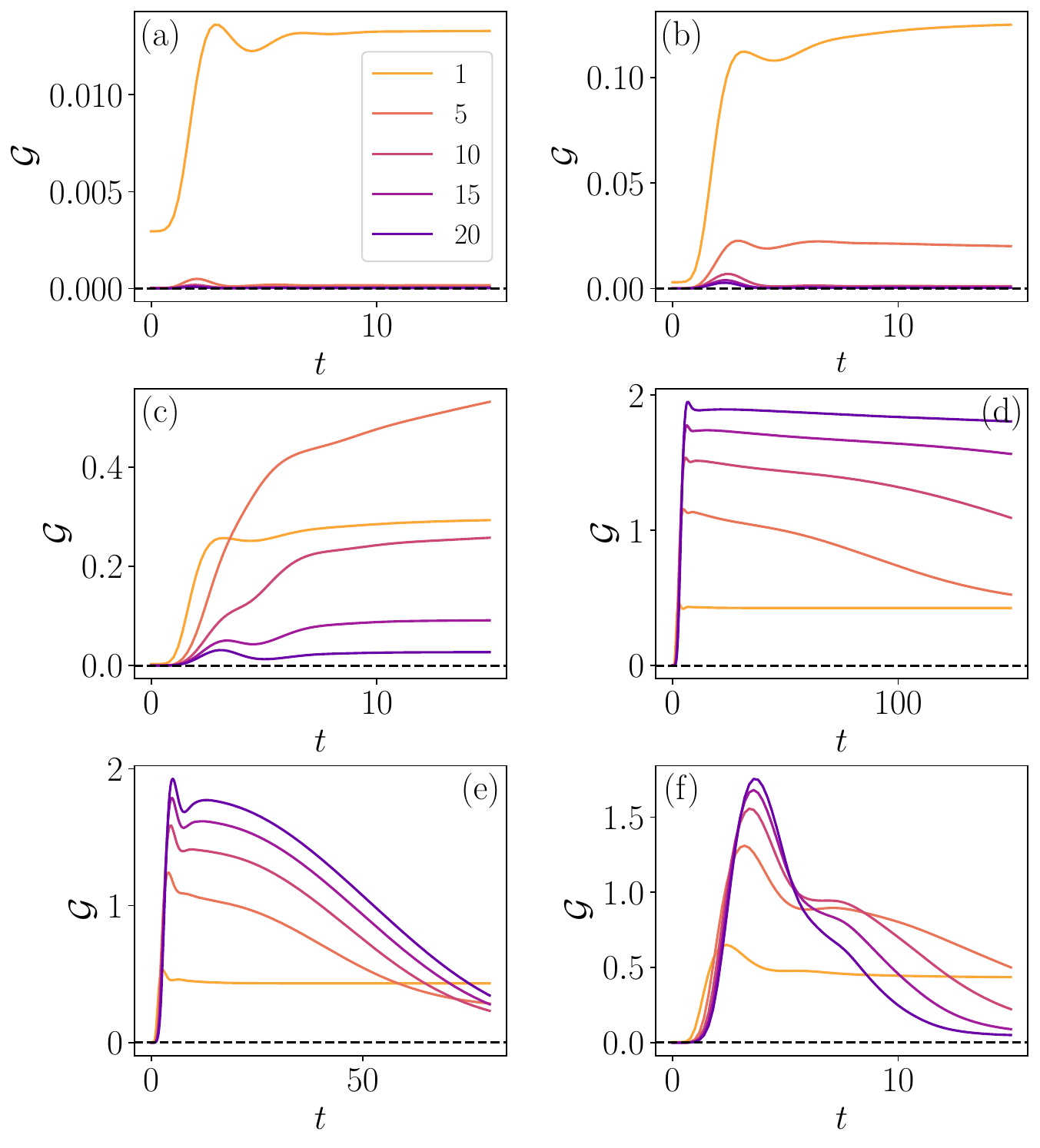}\quad
\caption{Degree of non-Gaussianity  [Eq.~\eqref{eq:gaussian_measure}], as a function of time, for the different quench protocols in Fig.~\ref{fig:firstmoment_with_plotmarkers}. Each curve correspond to a different value of $N$, as shown in panel (a). Other parameters are the same as Fig.~\eqref{fig:firstmoment_with_plotmarkers}.}
\label{fig:gaussian_measure}
\end{figure}

We can quantify the degree of non-Gaussianity in $\rho_t$ by comparing it with a corresponding Gaussian state $\rho_t^G$ having the same first and second moments. 
Following~\cite{genoni_quantifying_2010}, we do this using the quantum Kullback-Leibler divergence
 \begin{equation}
 \label{eq:gaussian_measure}
 \begin{split}
    \mathcal{G}(\rho_t) &=D(\rho_t || \rho_t^G) = \tr\Big\{ \rho_t \ln \rho_t - \rho_t \ln \rho_t^G\Big\}
    %= S(\rho_t^G)-S(\rho_t)\\ %= \tr{\rho_t \ln{\rho_t}} - \tr{\rho_t \ln{\rho_t^G}}\\
%                &=h(\sqrt{\det{\Theta}}) - S(\rho_t)
 \end{split}
 \end{equation}
 %
 %where $S(\rho_t) =  -\text{tr}(\rho\ln{\rho})$ is the von Neumann entropy.
 %and $\Theta$ is the covariance matrix, with elements $\Theta_{ij}=1/2\mean{\{r_i,r_j\}}-\mean{r_i}\mean{r_k}$, where the vector $r=(q,p)$ is composed of the quadrature operators $q=(\dg{a}+a)/\sqrt{2}$ and $p=i(\dg{a}-a)/\sqrt{2}$. Finally, we have the function $h(x)=(x+1/2)\ln{(x+1/2)}-(x-1/2)\ln{(x-1/2)}$. 
The state is Gaussian if $\mathcal{G} = 0$ and non-Gaussian otherwise. The results for $\mathcal{G}(\rho_t)$, for the quenches laid out in Fig.~\ref{fig:firstmoment_with_plotmarkers}, are presented in Fig.~\ref{fig:gaussian_measure}.

The quench in figure~\ref{fig:gaussian_measure}(a) goes to $\epsilon_f=0.6 < \epsilon_-$ and thus remains in the dark phase, outside the bistability region. We observe that most evolutions remain nearly Gaussian, except $N=1$. 
This is consistent with the fact that Gaussianity, even in the NESS, is only expected in the thermodynamic limit. 
In figure~\ref{fig:gaussian_measure}(b) the quench goes to $\epsilon_f = 0.8$. Again, it is still in the dark phase, but now it lies inside the bistability region and below  the critical point; that is,  $\epsilon_- < \epsilon_f < \epsilon_c$. Now we observe that  $N=5$ is not even close to being Gaussian, but one does roughly find Gaussianity for   $N=10$ onward.

The real action starts in Fig.~\ref{fig:gaussian_measure}(c), where $\epsilon_f=\epsilon_c$. 
The corresponding final NESS will  be roughly a mixture of  the dark and bright phases. We observe that the dynamics is markedly  non-Gaussian for any value of $N$. 
But for larger values of $N$ one still finds that there is a tendency toward becoming Gaussian.

The situation changes completely in figure~\ref{fig:gaussian_measure}(d), where we show results for $\epsilon_f=1.1$, thus representing a quench from the dark to the bright phase, but within the bistability region, $\epsilon_c < \epsilon_f < \epsilon_-$ (see Fig.~\ref{fig:firstmoment_with_plotmarkers}).
In this case we find that $\mathcal{G}(\rho_t)$ \emph{increases} with $N$.
This indicates that, in this case, a Gaussian limit would never be reached, regardless of the values of $N$ used. 
It therefore represents the most dramatic example, over all quenches, of a markedly non-Gaussian dynamics. 
A similar behavior is observed in Fig.~\ref{fig:gaussian_measure}(e), where $\epsilon_f = \epsilon_+$.
Unlike (d), however, we now see that $\mathcal{G}(\rho_t)$ tends to fall for long times. 
This is related to the time-scales of the problem, which are much slower in Fig.~\ref{fig:gaussian_measure}(d).
 
Finally, in figure~\ref{fig:gaussian_measure}(f) we show results for $\epsilon_f = 1.3 > \epsilon_+$.
It is found that, for intermediate times, the state is non-Gaussian for any size  $N$, but quickly tends back toward zero (notice the different horizontal scale in comparison with (d) and (e)). 

The above results, therefore, clearly show that during the quench dynamics, the state will in general be highly non-Gaussian. 
This justifies the use of the full numerical algorithm described above, in contrast to, say, Gaussianization techniques. 
It also highlights one of the advantages of using the Husimi function, which is non-negative for any quantum state (in contrast, for instance, with the Wigner function which can become negative). 

 %%%%%%%%%%%%%%%%%%%%%%%%%%%%%%%%%%%%%
 %
 %      ENTROPY FLUX DYNAMICS
 %
 %%%%%%%%%%%%%%%%%%%%%%%%%%%%%%%%%%%%%

\subsection{Entropy flux rate dynamics}
\label{Sec:Entropy_flux_dyn}

\begin{figure}[!t]
\centering
\includegraphics[width = 0.45\textwidth]{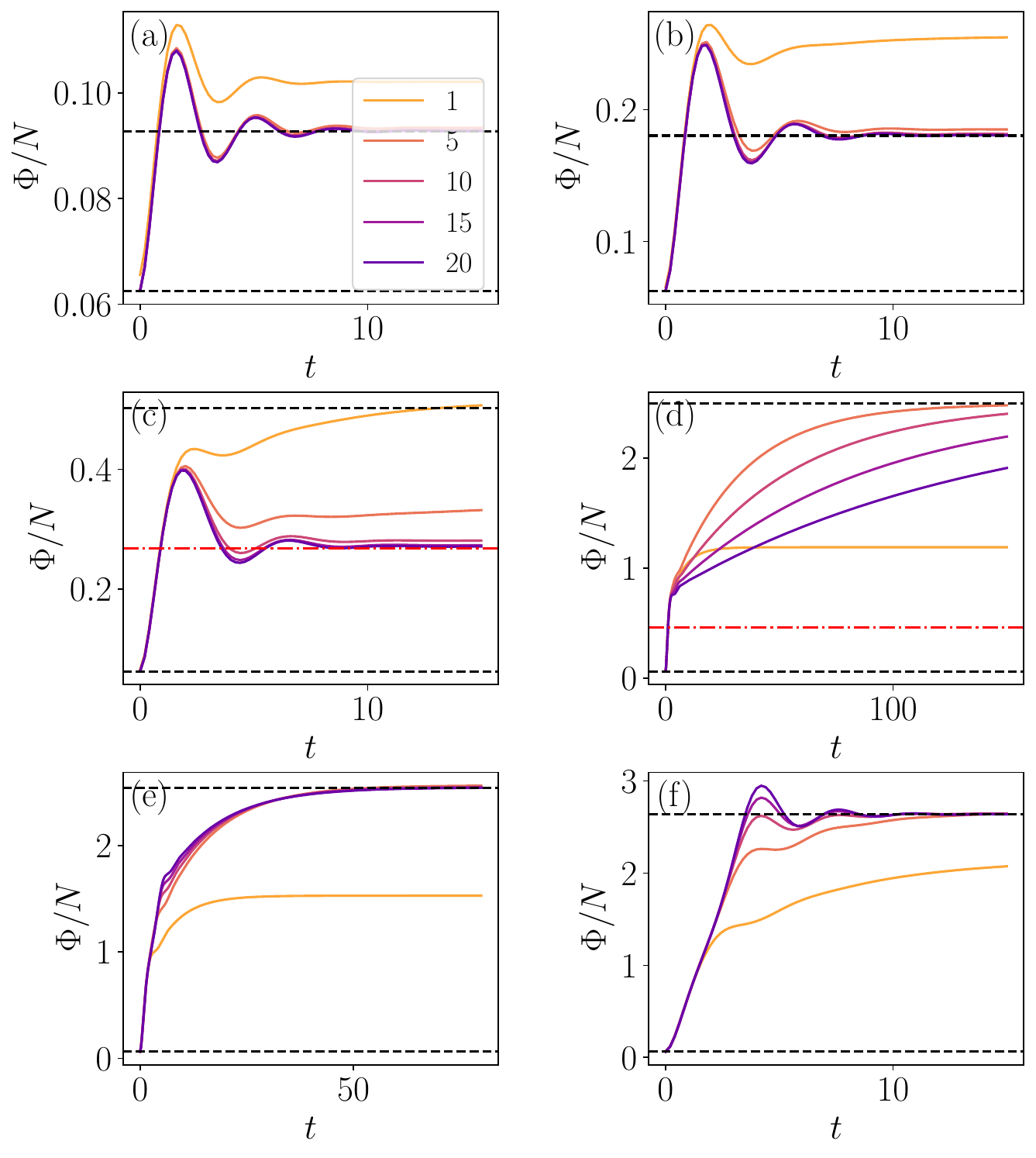}
\caption{$\Phi/N$, Eq.~\eqref{eq:Total_entropy_flux_rate}, as a function of time. The dashed black lines correspond to the NESS values at $\epsilon_i$ and $\epsilon_f$ (in the thermodynamic limit). \textcolor{black}{The red dash-dotted lines represent the MFA solution.}
Other details are as in Fig.~\ref{fig:gaussian_measure}.
}
\label{fig:Phi_T_quench}
\end{figure}

Fig. \ref{fig:Phi_T_quench} summarizes the results for the entropy flux rate~\eqref{eq:Total_entropy_flux_rate}.
This quantity is dominated by the first term, which is extensive in $N$. The dashed black lines represent the values of $\Phi_{\text{NESS}}$ computed from the exact solution~\eqref{eq:KBM-Exact_moments}. In Figs.\ref{fig:Phi_T_quench}(c) and \ref{fig:Phi_T_quench}(d), the MFA solution, Eq.~\eqref{eq:KBM_MFA_sol}, is represented by red dash-dotted lines, for comparison.

In Figs.~\ref{fig:Phi_T_quench}(a) and~\ref{fig:Phi_T_quench}(b) the system relaxes to the exact NESS after a relatively short transient (when contrasted to (d)-(f)). This is related to the fact that both the initial and final NESSs belong to the same manifold of dark-phase states. 
The behavior of the flux is in agreement with the non-Gaussianity in Fig.~\ref{fig:gaussian_measure}.
In Fig.~\ref{fig:Phi_T_quench}(c), where the quench goes to the critical point $\epsilon_c$,  the system tends to relax to the NESS predicted by the MFA solution in Eq.~\eqref{eq:KBM_MFA_sol} when $N\rightarrow \infty$. Recall that this quench was the first to present a non-Gaussian dynamics. 

In Fig.\ref{fig:Phi_T_quench}(d), where the quench is toward the bright phase, we begin to observe a non-trivial dependence on $N$ (all plots refer to $\Phi/N$). 
Recall that, in this case, the degree of non-Gaussianity increases with $N$. 
One also notices that in this case the system does not relax to the  exact NESS, nor to the MFA solution, even for the significantly longer simulation times (compare the horizontal axis with the other curves).  
Given a sufficient amount of time, the system would likely relax. However, this  requires significantly longer simulations, which are beyond the numerical capabilities of our code.

%The quench to the bright phase, inside the bistable region, Fig.\ref{fig:Phi_T_quench}(d), we observe a dependency on $N$ even for the scaled entropy flux. Recall that as $N$ the greater the $N$ the more non-gaussian the dynamics is, we also note that in this quench the system does not relax nor to the exact NESS neither the MFA solution of the NESS, although it seems to approach the MFA solution as $N$ increases.

Finally, in Figs.\ref{fig:Phi_T_quench}(e) and (f), one  observes a much more orderly transition between the two NESSs, as depicted by the dashed black lines. The transient time scales also tends to be relatively $N$-independent, except for $N = 1$.

\begin{figure}[!t]
\centering
\includegraphics[width = 0.45\textwidth]{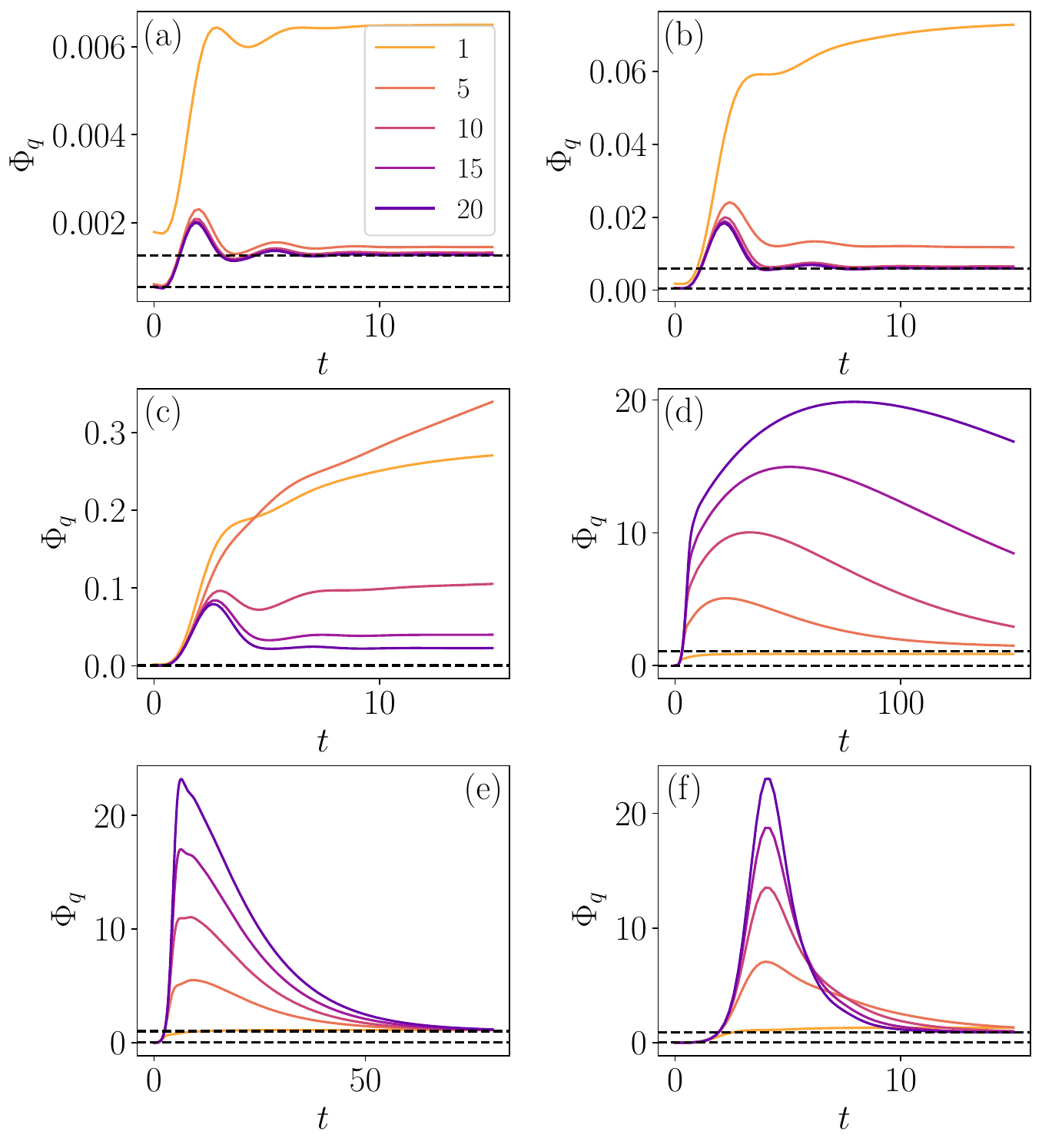}
\caption{$\Phi_q$, Eq.~\eqref{eq:splitting_entropy_flux_rate}, as a function of time. Other details are  as in Fig.~\ref{fig:gaussian_measure}.
}
\label{fig:Phiq_quench}
\end{figure}

In  Fig.~\ref{fig:Phiq_quench} we present  the plots for the quantum entropy flux $\Phi_q$ in Eq.~\eqref{eq:splitting_entropy_flux_rate}.
This quantity behaves similarly to $\Pi_d$, which will be discussed in the next section.
Thus, we shall comment more thoroughly on it below. 
At this point, we just call attention to the overall vertical scale of $\Phi_q$, in comparison with $\Phi$ in \textcolor{black}{Fig.~\ref{fig:Phi_T_quench}}. 
The latter is normalized by $N$. 
Notwithstanding, for certain quenches one finds that $\Phi_q$ can reach values that are comparable in magnitude to $\Phi_\text{ext}$. 
This is particularly true for $\epsilon_f > \epsilon_c$ (Figs.~\ref{fig:Phiq_quench}(d)-(f)).

 %%%%%%%%%%%%%%%%%%%%%%%%%%%%%%%%%%%%%
 %
 %      Q ENTROPY PRODUCTION DYN
 %
 %%%%%%%%%%%%%%%%%%%%%%%%%%%%%%%%%%%%%

\subsection{Quantum entropy production dynamics}

\begin{figure}[t]
\centering
\includegraphics[width=0.45\textwidth]{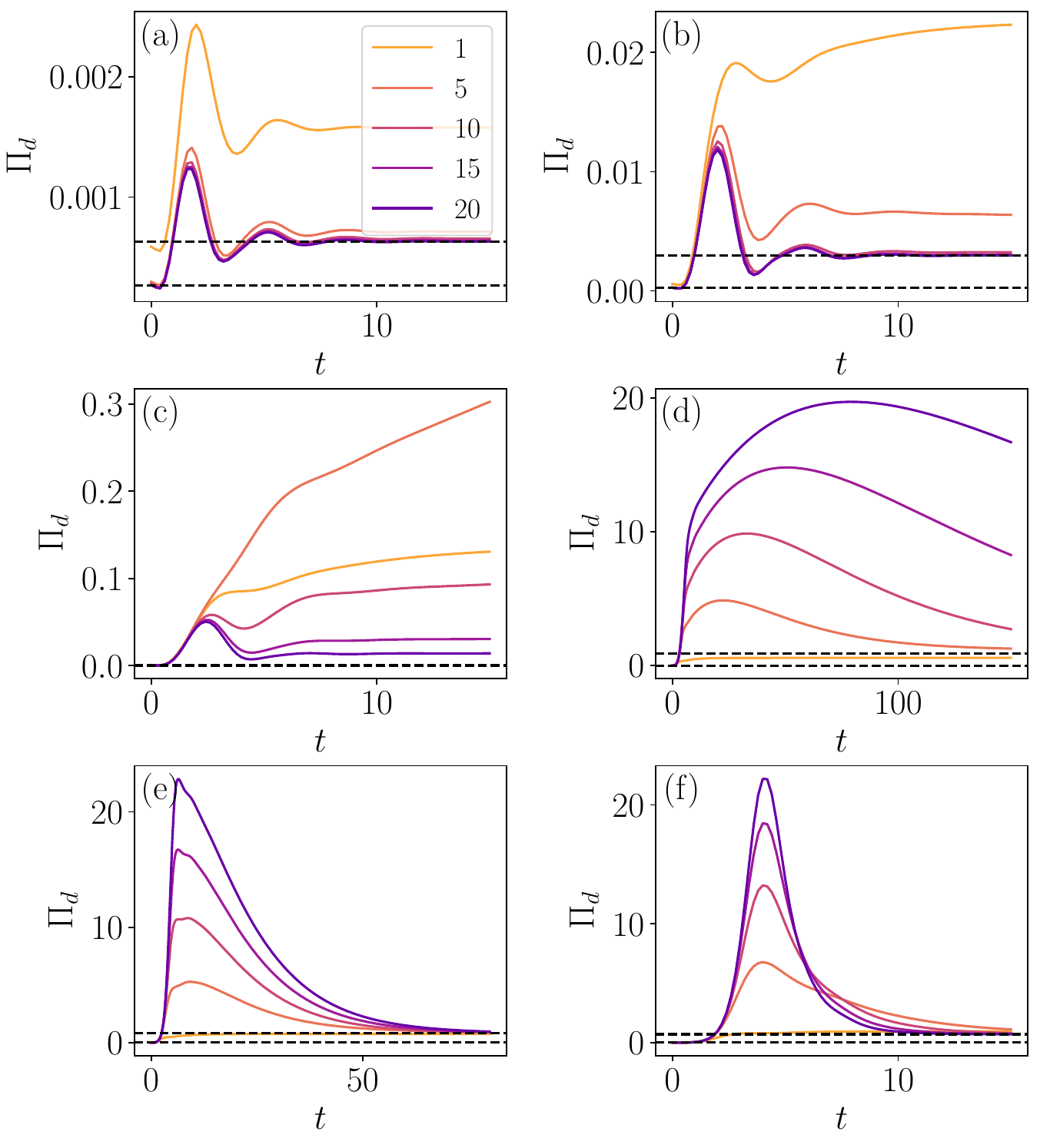}
\caption{
$\Pi_d$, Eq.~\eqref{eq:Pi_d}, as a function of time. 
The black dashed lines correspond to the NESS values in the thermodynamic limit (the line in image (c) is not shown since it falls outside the scale). Other details are as in Fig.~\ref{fig:gaussian_measure}.
}
\label{fig:Pies1_quench}
\end{figure}
\begin{figure}[t]
\centering
\includegraphics[width =0.45\textwidth]{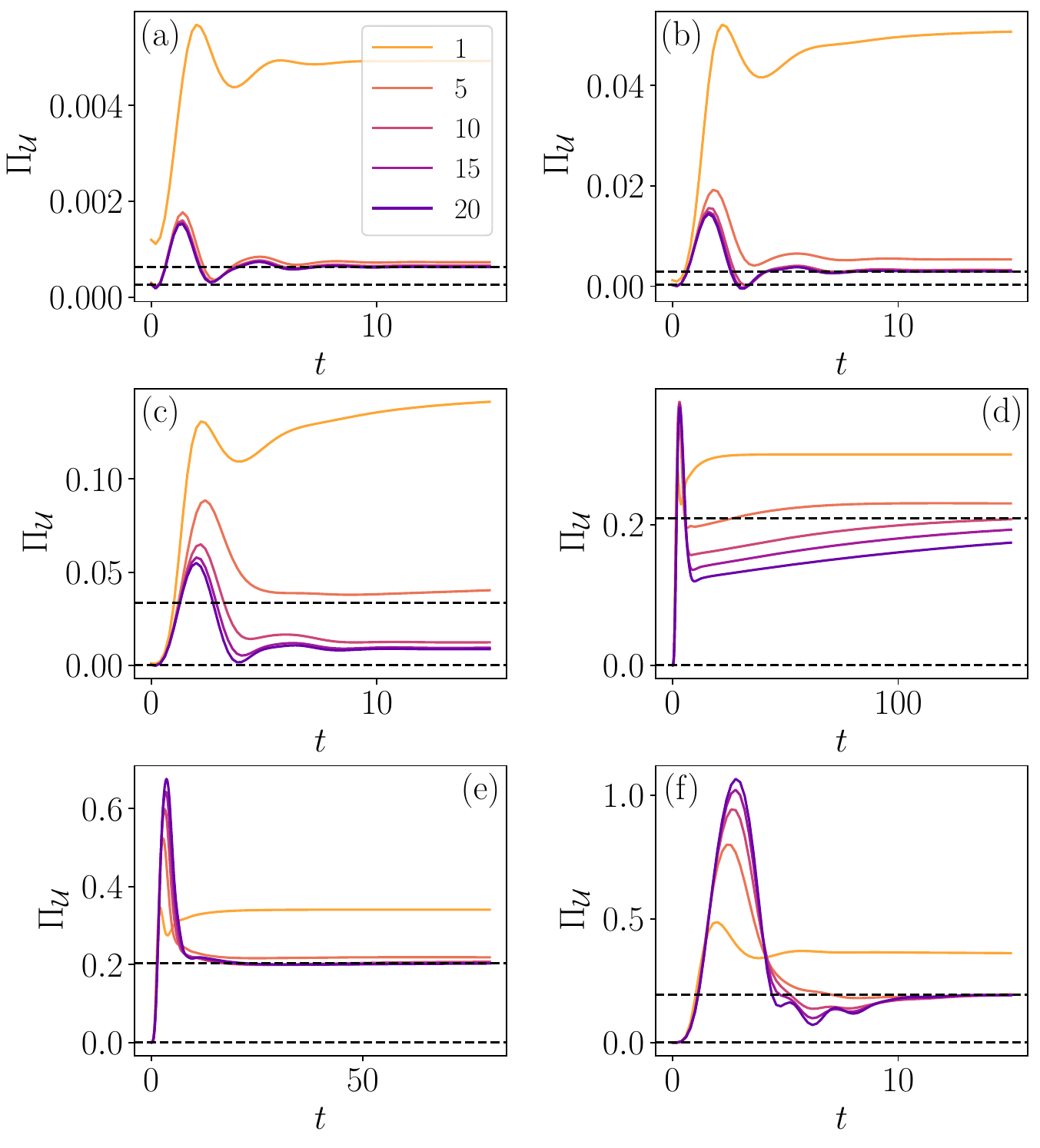}
\caption{$\Pi_\mathcal{U}$, Eq.~\eqref{eq:Pi_U_contribution}, as a function of time. 
Other details are as in Fig.~\ref{fig:gaussian_measure}.
}
\label{fig:Pies2_quench}
\end{figure}

In this section we present and compare the dynamics of the quantum entropy production rate due to dissipation, $\Pi_d$ ( Fig.~\ref{fig:Pies1_quench}), and due to the  unitary contribution, $\Pi_{\mathcal{U}}$ (in Fig.~\ref{fig:Pies2_quench}). 
We begin by comparing the two quantities in the cases of  (a) $\epsilon_f = 0.6$, (b) $\epsilon_f = 0.8$ and (c) $\epsilon_f = \epsilon_c$. 
In these three cases, we observe that $\Pi_d$ and  $\Pi_{\mathcal{U}}$ are of the same order of magnitude. Physically, this means that irreversibility due to dissipation is comparable to the one introduced by the coarse-graining. 
We call attention to a non-trivial  $N$-dependency in $\Pi_d$ in Fig.~\ref{fig:Pies1_quench}(c), where the curve for $N= 5$ is actually above $N=1$, unlike all other cases. 
It is not clear to us, precisely why this happens.

% remember that from this quench onward the dynamics is non-gaussian (see Fig.\ref{fig:gaussian_measure}(c)-(f)) and a relaxation to the MFA prediction of the NESS (Fig.\ref{fig:Phi_T_quench}(c)).
%In Fig.~\ref{fig:Pies1_quench} we show the dynamics of the quantum entropy production due to dissipation $\Pi_d$ in the upper row and the unitary contribution $\Pi_u$ in the lower row. First, we observe that for the quenches to $\epsilon_f = 0.6$ and $\epsilon_f = 0.8$, for which, as we have seen in Sec.~\ref{sec:Gauss_tests}, the dynamics is effectively gaussian. The magnitude of $\Pi_d$, \ref{fig:Pies1_quench}(a), \ref{fig:Pies1_quench}(c) and \ref{fig:Pies1_quench}(e), is comparable with the magnitude of $\Pi_u$, \ref{fig:Pies1_quench}(b), \ref{fig:Pies1_quench}(d) and \ref{fig:Pies1_quench}(f). \textcolor{black}{Physically, it mean that irreversibility due to dissipation is indistinguishable from the one introduced by the heterodyning procedure, i.e. the loss of information due to performing a measurement in the coherent state basis. THINK ABOUT HOW TO DISCUSS THE BEHAVIOR IN (E) FOR $N=5$ ->}

Conversely, for the quenches to (d) $\epsilon_f = 1.1$, (e) $\epsilon_f = \epsilon_+$ and (f) $\epsilon_f = 1.3$,
we find that $\Pi_d \gg \Pi_\mathcal{U}$; 
that is, dissipation becomes the main source of irreversibility. In all these cases both $\Pi_d$ and $\Pi_{\mathcal{U}}$ present an $N$-dependent transient, peak at some instant of time and then eventually relaxing back to the NESS value. 
Notice how the maximal values reached by $\Pi_d$ are significantly higher than those of the NESS.
This allows one to conclude that the transition from a dark to a bright NESS is accompanied by a significant production of entropy. 
Please note also the horizontal scales, showing that the relaxation times in Figs.~\ref{fig:Pies1_quench}(d) and \ref{fig:Pies2_quench}(d) are significantly larger. 
Conversely, in Figs.~\ref{fig:Pies1_quench}(f) and \ref{fig:Pies2_quench}(f), the relaxation times reduce significantly. Notwithstanding, the maximum entropy production rate is still very high.

%{Physically, we have a system that is initially in the NESS state living in the "manifold" $\mathcal{M}_1$ of the dark phase, by making a quench we are abruptly taking the system out of this manifold, on the top of that we want the system to go a different manifold $\mathcal{M}_2$ of the bright phase. The "path" the system has to go in order to do so is longer, what is pinpointed by the bigger timescale involved $t=150$. But in the middle of the way, it find a shortcut, the bistable NESS manifold described by the "union" of both phase $\mathcal{M}_{\text{bi}} =\mathcal{M}_1 \cup \mathcal{M}_2$, it relax to this state them. Also, as the system starts receiving much more photons than it had initially, the dissipator has to "work" more to throw them out, thus leading to more dissipation, hence a more irreversible process.}The most drastic case is shown in Fig.\ref{fig:Pies1_quench}(d) where the transient is long lived, the greater the $N$ the longer the system stays in the peak of $\Pi_d$. The peak for $\Pi_{\mathcal{U}}$ happens in the same instant for every $N$ as we observe from Fig.\ref{fig:Pies2_quench}(d). 

% \textcolor{black}{Next, in Figs.~\ref{fig:Pies1_quench}(e) and \ref{fig:Pies2_quench}(e), the quench is to $\epsilon_f = \epsilon_+$. As in the case just discussed we have a rapidly growing of the entropy production $\Pi_d$ at transient times but here the dissipator is able to take the system assymptotically to the exact NESS for long relaxation times.}

These results show that the thermodynamic response of the system depends sensibly on the final pump value $\epsilon_f$.
When the quench is to the same phase we find that $\Pi_d \approx \Pi_{\mathcal{U}}$, while when we abruptly change the phase $\Pi_d \gg \Pi_{\mathcal{U}}$: in the former, the contributions of the dissipation and uncertainty introduced by the coarse-graining are comparable, and in the latter the dissipation dominates over {\color{black}coarse graining that produces $\Pi_{\mathcal{U}}$}. When the system is quenched toward the bright phase, one finds a strong dependence on $N$, together with a peak of the entropy production rates before the relaxation to the NESS. 
Whether $\epsilon_f$ lies inside or outside the bistability region, plays an essential role in the time-scales involved, which become significantly slower due to the metastable character of this region.

%%%%%%%%%%%%%%%%%%%%%%%%%%%%%%%%%%%%%%%%%%%%%%%%%
%
%
%               CONCLUSION
%
%
%%%%%%%%%%%%%%%%%%%%%%%%%%%%%%%%%%%%%%%%%%%%%%%%%

\section{Conclusion and outlook}
\label{Sec:conclusions}

Entropy production is the key concept characterizing systems out of equilibrium. 
And much is still unknown about how it behaves as a system undergoes a non-equilibrium transition. 
In this paper we have employed a phase-space formalism to characterize the entropy production of the Kerr bistability model, a prototypical model of driven-dissipative quantum optical systems, presenting a discontinuous transition. 
This paper complements Ref.~\cite{goes_quantum_2020}, which studied the NESS of this model. 
Here our focus was on dynamical aspects of the entropy production in a sudden quench scenario.

We showed that, in general, the state during the dynamics is highly non-Gaussian. By analyzing the quenches to different representative configurations, we have found that the thermodynamic response of the system presents at least three markedly different behaviors. 
For quenches within the same phase, the unitary and dissipative contributions to the entropy production are of the same order of magnitude and the time-scales involved are all relatively short and $N$-independent. 
For quenches from the dark to the bright phase, but still inside the bistability region, we find that $\Pi_d \gg \Pi_\mathcal{U}$ and the time-scales become significantly longer. 
Finally, for quenches from the dark to the bright phase, but outside the bistability region, one still has $\Pi_d \gg \Pi_\mathcal{U}$, but now the time-scales become once again relatively short and $N$-independent. 

By gauging how far the system is from equilibrium, together with the relative contribution from unitary and dissipative dynamics, this study has therefore helped shed light on the intricate interplay between criticality and dissipation in non-equilibrium transitions. 
It also serves to tighten the connection between quantum optical models and statistical mechanics. 

Finally, we believe that these results also show the flexibility of the Wehrl entropy production framework in dealing with driven-dissipative quantum optical systems far from equilibrium. 
In future research, we plan to consider infinitesimal quenches, as well as models presenting continuous transitions. 
From a computational point of view, these studies would benefit tremendously from more efficient methods of computing the Husimi $Q$-function and the associated phase space quantities. 
This would be an absolute necessity if one were to extend these studies, for instance, to multimode systems such as the Dicke model or the optical parametric oscillators. 

\emph{Acknowledgements - }
The authors acknowledge fruitful discussions with C. Fiore and J. Goold. 
G. T. L. acknowledges the financial support of the S\~ao Paulo Funding Agency FAPESP (Grants No. 2017/50304-7, 2017/07973-5 and 2018/12813-0) and the Brazilian funding \textcolor{black}{agency CNPq (Grant No. INCT-IQ 246569/2014-0). B. O. G. acknowledges the financial support from the Brazilian funding agencies CNPq and CAPES}, D. M. Valente, S.V. Moreira and J. P. Santos for insightful comments on the work.

%%%%%%%%%%%%%%%%%%%%%%%%%%%%%%%%%%%%%%%%%%%%%%%%%
%
%
%               APPENDICES
%
%
%%%%%%%%%%%%%%%%%%%%%%%%%%%%%%%%%%%%%%%%%%%%%%%%%

\appendix
\label{Appendix}

%%%%%%%%%%%%%%%%%%%%%%%%%%%%%%%%%%%%%
 %
 %      HETERODYNE MEASUREMENT
 %
 %%%%%%%%%%%%%%%%%%%%%%%%%%%%%%%%%%%%%
 
 \section{Heterodyne measurements}
\label{App:heterodyne_mesurement}

In order to clarify the physical content of the Wehrl entropy, Eq.~\eqref{eq:Wehrl_entropy_def}, we briefly review here the interpretation of the Husimi $Q$-function  as the probability distribution of the outcomes of a heterodyne measurement. We start by considering a generalized measurement, described by the continuous set of Kraus operators
\begin{equation}
    M_{\mu} = \frac{1}{\sqrt{\pi}}\ket{\mu}\bra{\mu}.
\end{equation}
This set is properly normalized, which is a consequence of the (over)completeness of the coherent states basis,
% %
% \begin{equation*}
%     \frac{1}{\pi}\int \dd^2\mu \;\ket{\mu}\bra{\mu}=1,
% \end{equation*}
% %
% it is obvious that this set is properly normalized,
% %
\begin{equation}
    \int \dd^2\mu \;\dg{M}_{\mu} M_{\mu}=\int \dd^2\mu \;\;\frac{\ket{\mu}\bra{\mu}}{\pi} =1.
\end{equation}
Thus,  the probability of obtaining outcome $\mu$ will be given by,
\begin{equation}
    p_{\mu}=\tr{M_{\mu}\rho \dg{M}_{\mu}} = \frac{1}{\pi}\bra{\mu}\rho\ket{\mu} \equiv Q(\mu,\cj{\mu}),
\end{equation}
which is precisely the Husimi $Q$-function.
The Wehrl entropy~\eqref{eq:Wehrl_entropy_def} can therefore be interpreted in an operational sense, as the entropy associated with the outcome probability distribution of an heterodyne measurement. 
This highlights its coarse-grained character, as it also encompasses the additional noise introduced by the measurement.

\section{Computation of $\Pi_J$ and $\Pi_{\mathcal{U}}$ from Bargmann state}
\label{App:Details_Pies}

To compute $\Pi_J$ and $\Pi_{\mathcal{U}}$ numerically, we make use of the Bargmann state, defined as,
\begin{equation}
\label{eq:def_Barg_state}
    ||\mu\rangle = \exp{\frac{|\mu|^2}{2}}\ket{\mu}.
\end{equation}
where $\ket{\mu}$ is the usual coherent state. This state has the useful property that the action of the derivatives with respect $\mu$ and $\cj{\mu}$ is mapped into the action of the creation/annihilation operators as,
\begin{equation}
\label{def:Barg_derivatives}
    \begin{split}
    \pd{\mu}{||\mu\rangle} &= \dg{a}||\mu\rangle\\
    \pd{\cj{\mu}}{\langle\mu||} &= \langle\mu||a
    \end{split}
\end{equation}
We can use this to write the phase-space currents~\eqref{J} in a computationally more convenient way. 
First we express the $Q$-function as
\begin{equation}
    Q(\mu,\cj{\mu}) = \frac{\exp{-|\mu|^2}}{\pi}\langle\mu||\rho||\mu\rangle.
\end{equation}
Using Eq.~\eqref{def:Barg_derivatives} to compute $\pd{\cj{\mu}}{Q}$, we then find
\begin{equation}
\label{eq:1st_Barg_derivative}
\begin{split}
    \pd{\cj{\mu}}{Q} &= - \mu Q  +\frac{\exp{-|\mu|^2}}{\pi}\langle\mu||a\rho||\mu\rangle\\
    &= - \mu Q  +\frac{1}{\pi}\langle\mu|a\rho|\mu\rangle,
\end{split}
\end{equation}
where, in the second line, we already recast the state in terms of usual coherent states. 
We then find that
\begin{equation}
    J(Q) = - \frac{\kappa}{\pi}\langle\mu|a\rho|\mu\rangle
\end{equation}
which allows us to write $\Pi_J$ in Eq.~\eqref{eq:entropy_production_rate} as,
\begin{equation}
    \Pi_J = \frac{2\kappa}{\pi}\int \dd^2\mu\;\frac{|\langle\mu|a\rho|\mu\rangle|^2}{\langle \mu |\rho | \mu \rangle}.
\end{equation}
Now, we turn to the computation of Eq.~\eqref{eq:Pi_U_contribution}. First, we note it can be rewritten as,
\begin{equation}
    \Pi_{\mathcal{U}} = U \int \dd^2\mu\; \Im{\frac{\cj{\mu}^2(\pd{\cj{\mu}}{Q})^2}{Q}}.
\end{equation}
Hence, using Eq.~\eqref{eq:1st_Barg_derivative} and the definition of $Q(\mu,\cj{\mu})$, we obtain
\begin{equation*}
    \Pi_{\mathcal{U}} = U \int \dd^2\mu\; \Im{\frac{|\mu|^2}{\pi}\bigpar{|\mu|^2\bra{\mu}\rho\ket{\mu} + 2\cj{\mu}\bra{\mu} a\rho\ket{\mu}} + \cj{\mu}\frac{\bra{\mu}a\rho\ket{\mu}^2}{\bra{\mu}\rho\ket{\mu}}}
\end{equation*}
During the simulations, the quantities $\langle \mu |\rho | \mu \rangle$ and $\langle\mu|a\rho|\mu\rangle$ are  directly computed from $\rho_t$.

\bibliography{references}

\end{document}